\documentclass[12pt]{article}
\usepackage{fullpage}
\usepackage{amsmath}
\usepackage{amssymb}
\usepackage[dvips]{epsfig}

\def\##1{\underline{#1}}
\def\=#1{\underline{\underline{#1}}}

\def\+#1{\underline{\bf #1}}
\def\*#1{\underline{\underline{\bf #1}}}

\def\r#1{(\ref{#1})}
\def\l#1{\label{#1}}
\def\c#1{\cite{#1}}

\def\le{\left(}
\def\ri{\right)}
\def\les{\left[}
\def\ris{\right]}
\def\lec{\left\{}
\def\ric{\right\}}

\def\.{\mbox{ \tiny{$^\bullet$} }}

\def\epso{\epsilon_{\scriptscriptstyle 0}}

\def\muo{\mu_{\scriptscriptstyle 0}}

\def\eps{\epsilon}

\begin{document}

\begin{center}

{\bf {\Large Uniaxial dielectric mediums with hyperbolic
dispersion relations }}

 \vspace{10mm} \large

Tom G. Mackay\footnote{Corresponding author. E--mail: T.Mackay@ed.ac.uk}\\
{\em School of Mathematics,
University of Edinburgh, Edinburgh EH9 3JZ, UK}\\
\bigskip
 Akhlesh  Lakhtakia\footnote{E--mail: akhlesh@psu.edu; also
 affiliated with Department of Physics, Imperial College, London SW7 2 BZ, UK}\\
 {\em CATMAS~---~Computational \& Theoretical
Materials Sciences Group\\ Department of Engineering Science and
Mechanics\\ Pennsylvania State University, University Park, PA
16802--6812, USA}\\
\bigskip
Ricardo A. Depine \footnote{ E--mail: rdep@df.uba.ar}\\
{\em Grupo de Electromagnetismo Aplicado, Departamento de
Fisica,\\
Facultad de Ciencias Exactas y Naturales, Universidad de Buenos
Aires, Ciudad Universitaria, Pabell\'{o}n I,
1428 Buenos Aires, Argentina}\\

\end{center}

\vspace{4mm}

\normalsize

\begin{abstract}
The dispersion relations for conventional uniaxial dielectric
mediums may be characterized as elliptical or elliptical--like,
according to whether the medium is nondissipative or dissipative,
respectively. However, under certain constitutive parameter
regimes, the dispersion relations may be hyperbolic or
hyperbolic--like. We investigate planewave propagation in a
hyperbolic/hyperbolic--like uniaxial dielectric medium. For both
evanescent and nonevanescent propagation, the phase velocity is
found to be positive with respect to the time--averaged Poynting
vector. A conceptualization of a hyperbolic--like uniaxial medium
as a homogenized composite medium is presented.
\end{abstract}


\section{Introduction}

 As the materials sciences and technologies continue their rapid
development, realistic possibilities are emerging of realizing
so--called \emph{metamaterials} with  novel and hitherto
unconsidered optical/electromagnetic properties. A prime example
is provided by the recently discovered metamaterials which support
planewave propagation with \emph{negative phase
 velocity} (NPV), and thereby negative refraction. Until 2000,
  little attention had been paid to the
 phenomenon of negative refraction.
Since 2000, there has been an explosion of interest in negative
refraction \c{Pendry04,SAR}, following experimental reports of a
metamaterial which supports negative refraction in the microwave
regime \c{SSS}.

 Naturally--occurring uniaxial crystals have  been extensively
studied ever since the earliest days of the
 optical  sciences. However, the electromagnetic properties
  of uniaxial mediums have recently been revisited by theoreticians in
  consideration of the prospects for
   NPV propagation in such mediums \c{Kark,Liu,Yonghua,Perez}. A closely related issue concerns
    uniaxial dielectric--magnetic mediums with indefinite
constitutive dyadics \c{Smith03,Smith04}.

The defining characteristic of a uniaxial dielectric medium is a
distinguished axis of symmetry, known as the optic axis.
Mathematically, the permittivity dyadic of  a uniaxial dielectric
medium may be expressed as
\begin{equation}
\=\eps = \eps \,\=I + \le \eps_x - \eps \ri \hat{\#x}
\,\hat{\#x}\,, \l{perm_dyadic}
\end{equation}
where a coordinate system has been selected in which the direction
of the optic axis  coincides with the direction of the unit vector
$\hat{\#x}$ lying along the $x$ axis, and $\=I$ denotes the
3$\times$3 identity dyadic. The real--valued parameter
\begin{equation}
\gamma = \left\{
\begin{array}{lcr}
\displaystyle{\frac{\eps_x}{\eps}} & \mbox{for} & \eps_x, \eps \in
\mathbb{R}
\\ &&
\\
\displaystyle{\frac{\mbox{Re} \lec \eps_x \ric}{\mbox{Re} \lec
\eps \ric }} & \mbox{for} & \eps_x, \eps \in \mathbb{C}
\end{array}
\right.  \l{gamma}
\end{equation}
may be usefully employed to characterize planewave propagation in
the medium specified by \r{perm_dyadic}. The upper expression is
appropriate to nondissipative mediums whereas the lower expression
is appropriate to dissipative mediums.

The electromagnetic/optical properties of uniaxial mediums with
$\gamma > 0$~---~this category includes naturally--occurring
uniaxial crystals~---~
   have long been
established. Comprehensive descriptions can be found in standard
works \c{BW,Chen}. Uniaxial mediums with
 $\gamma < 0$ are much more exotic. Interest in these mediums
 stems from their potential applications in negatively refracting
 scenarios \c{Smith03,Smith04}  and in diffraction gratings \c{DL05}, for example.

 Planewave
propagation in a uniaxial  medium is characterized in terms of a
dispersion relation  which is quadratic in terms of the
corresponding wavevector components.
 The  dispersion relations for nondissipative
mediums with $\gamma > 0$ have an elliptical representation,
whereas a hyperbolic representation is associated with $\gamma <
0$. In this communication we investigate the planewave
characteristics and conceptualization of uniaxial dielectric
mediums with hyperbolic dispersion relations.

\section{Planewave analysis}

\noindent The propagation of plane waves with field phasors
\begin{equation}
\left.\begin{array}{l}
\#E(\#r) = \#E_0\, \exp \le i \#k \. \#r  \ri  \\[5pt]
\#H(\#r) = \#H_0\, \exp \le i \#k \. \#r \ri
\end{array}\right\}
\l{pw}
\end{equation}
in the uniaxial dielectric medium specified by the  permittivity
dyadic \r{perm_dyadic} is investigated. The permittivity
parameters are generally complex--valued;
  i.e.,   $\eps, \eps_x \in \mathbb{C}$.
The wavevector $\#k$  is taken to be of the form
\begin{equation}
\#k = \alpha \, \hat{\#x} + \beta \, \hat{\#z}\,, \l{kab}
\end{equation}
where
  $\alpha \in \mathbb{R}$,  $\beta
\in \mathbb{C}$ and $\hat{\# z}$ is the unit vector directed along
the $z$ axis. This form of $\#k$ is appropriate to planar boundary
value problems \c{Chen} and from the practical viewpoint of
potential optical devices \c{DL05}. We note that the plane waves
\r{pw} are generally nonuniform.

The
 source--free Maxwell curl postulates
\begin{equation}
\left.\begin{array}{l}
\nabla \times \#E(\#r) = i \omega \#B ( \#r )\\[5pt]
\nabla \times \#H(\#r) = - i \omega \#D ( \#r )
\end{array}\right\}
\l{Maxwell}
\end{equation}
yield
 the vector Helmholtz equation
\begin{equation}
\les \,\le \nabla \times \=I \,\ri \. \le \nabla \times \=I\, \ri
- \muo \omega^2 \=\eps\,\ris \.  \#E (\#r) = \#0\,, \l{Helmholtz}
\end{equation}
with $\muo$ being the permeability of free space. Combining \r{pw}
with \r{Helmholtz} yields the planewave dispersion relation
\begin{equation}
\le \alpha^2 + \beta^2 - \eps \muo \omega^2 \ri \le \alpha^2
\eps_x + \beta^2 \eps - \eps_x \eps \muo \omega^2 \ri = 0\,.
\l{disp}
\end{equation}

In the following  we consider the  time--averaged
Poynting vector
\begin{equation}
\#P (\#r)  = \frac{ \exp \le -2 \,
\mbox{Im} \lec \beta \ric  z \ri  }{2 \muo \omega} \,
\mbox{Re} \lec \, | \#E_0 |^2 \,\#k^* - \le \#E_0 \. \#k^* \ri \#E^*_0
\, \ric\,.
\end{equation}
Evanescent plane waves are characterized by  $\mbox{Im} \lec \beta
\ric
> 0 $. The scenario characterized
by $\mbox{Im} \lec \beta \ric < 0 $ is not physically plausible
for passive mediums and is therefore not considered here.

\subsection{Ordinary wave}

The ordinary  wavevector
\begin{equation}
\#k_{or} = \alpha \, \hat{\#x} + \beta_{or} \, \hat{\#z}\,,
\end{equation}
 arises  from the dispersion relation
\r{disp} with components satisfying
\begin{eqnarray}
\alpha^2 + \beta^2_{or}
& =& \omega^2 \eps \muo  \,.\l{k_or}
\end{eqnarray}
The vector Helmholtz equation
\r{Helmholtz} yields
 the eigenvector solution $\#E_0 = E_y
\hat{\#y}$, directed parallel to the unit vector $\hat{\# y}$
lying along the $y$ axis,  where the complex--valued magnitude
$E_y$ is determined by the initial/boundary conditions.
Consequently, the time--averaged Poynting vector reduces to
\begin{equation}
\#P (\#r)  = \frac{ \exp \le -2 \, \mbox{Im} \lec \beta_{or} \ric  z \ri}{2 \omega
\muo} \, | E_y |^2 \, \mbox{Re} \lec \#k^*_{or} \ric \,.
\end{equation}
Since
\begin{eqnarray}
\mbox{Re} \lec \#k_{or} \ric \. \#P (\#r) & =& \frac{ \exp \le -2
\, \mbox{Im} \lec \beta_{or}
\ric z \ri }{2 \omega \muo} \,  | E_y |^2 \, \les \alpha^2 +  \le \mbox{Re} \lec
\beta_{or} \ric \ri^2
 \ris \,
\l{k_or_npv}
   \ge \, 0\,,
\end{eqnarray}
we say that  ordinary plane waves have positive phase velocity
(PPV) for all directions of propagation.

Let us focus attention on  a nondissipative medium (i.e., $\eps,
\eps_x \in \mathbb{R}$). From  \r{k_or} we see that $\mbox{Im}
\lec \beta_{or} \ric \neq 0 $ for (i) $\eps > 0$ when $\omega^2
\eps \muo < \alpha^2$; and (ii) $ \eps < 0$. Thus, nonevanescent
ordinary plane waves propagate in a nondissipative medium only
when  $\eps > 0$ and $ - \omega  \sqrt{ \eps \muo} < \alpha <
  \omega  \sqrt{ \eps \muo}$.
In geometric terms, the wavevector components  have a circular
representation in $(\alpha, \beta_{or})$ space.

\subsection{Extraordinary wave}

The extraordinary wavevector
\begin{equation}
\#k_{ex} = \alpha \, \hat{\#x} + \beta_{ex} \, \hat{\#z}\,,
\end{equation}
 arises  from the dispersion relation
\r{disp} with components satisfying
\begin{eqnarray}
\alpha^2  \eps_x + \beta^2_{ex} \eps  &= & \omega^2 \eps \,
\eps_x \muo  \l{k_ex}\,.
\end{eqnarray}
In the case where $\beta_{ex} = 0$ the mathematical description of
the extraordinary wave is isomorphic to that for the ordinary
wave. Therefore, we exclude this possibility from our
consideration in this section.
  The
eigenvector
\begin{equation}
\#E_0 = \le \hat{\#x}   -  \, \frac{\eps_x \, \alpha}{\eps \,
\beta_{ex}} \, \hat{\#z} \,\ri E_x ,
\end{equation}
arises as a solution to  the vector Helmholtz equation
\r{Helmholtz}; the complex--valued magnitude $E_{x}$ is determined
by the initial/boundary conditions. The corresponding
 time--averaged Poynting vector is
provided as
\begin{eqnarray}
\#P (\#r) &=&
 \frac{
\exp \le -2 \, \mbox{Im} \lec \beta_{ex} \ric  z \ri
 }{2 \omega
\muo} \,\mbox{Re} \Bigg\{ \alpha \le  \left| \frac{\eps_x}{\eps \,
\beta_{ex}}
 \right|^2  \alpha^2  +
 \frac{\eps_x \beta^*_{ex}}{ \eps \, \beta_{ex}} \,\ri
  \hat{\#x}
   \nonumber \\&& +
   \le \beta^*_{ex} +  \alpha^2
  \frac{\eps^*_x}
  {\eps^* \, \beta^*_{ex}} \, \ri \hat{\#z}\,
 \Bigg\} | E_x |^2\,.
\end{eqnarray}
Hence, we find
\begin{eqnarray}
\mbox{Re} \lec \#k_{ex} \ric  \. \#P (\#r)  &=& \frac{ \exp \le
-2 \, \mbox{Im} \lec
\beta_{ex} \ric  z \ri
 }{2 \omega
\muo } \, \Bigg[ \le \mbox{Re} \lec \beta_{ex} \ric \ri^2
\nonumber \\ && + \alpha^2 \le \alpha^2 \left| \frac{\eps_x }{\eps
\, \beta_{ex}} \right|^2 + \mbox{Re} \lec \frac{ \eps_x \,
\beta^*_{ex}}{ \eps \, \beta_{ex}} \ric + \mbox{Re} \lec
\beta_{ex} \ric \, \mbox{Re} \lec \frac{\eps^*_x}{\eps^* \,
\beta^*_{ex}} \ric \ri \Bigg] \,. \l{kP}
\end{eqnarray}

We analytically explore  the nondissipative scenario for
nonevanescent and evanescent planewave propagation in Sections~2.3
and 2.4, respectively, whereas both the dissipative and the
nondissipative scenarios are treated graphically in Section~2.5.

\subsection{Nonevanescent propagation}

By \r{k_ex}, the inequality
\begin{equation}
\omega^2 \eps_x \muo - \alpha^2 \gamma > 0 \l{nonevan}
\end{equation}
is satisfied for  nonevanescent planewave propagation in a
nondissipative medium, where $\gamma$ is defined in \r{gamma}.
 Thus, $\mbox{Im} \lec \beta_{ex}
\ric = 0 $.
 We explore
the cases $\gamma > 0$ and $\gamma < 0$ separately.
\begin{itemize}
\item[(i)] If $\gamma
> 0$ then we require $ - \omega \sqrt{\eps \muo} < \alpha <
 \omega \sqrt{\eps \muo}$ in order to comply with \r{nonevan}.
This implies that
 $\eps
> 0$ and $\eps_x
> 0$.
In geometric terms, the wavevector components have an elliptical
representation in  $(\alpha, \beta_{ex})$ space.
\item[(i)] If $\gamma
< 0$ then the inequality \r{nonevan}  reduces to $\omega^2 \eps
\muo < \alpha^2$. Therefore, we see that nonevanescent propagation
arises for (a) $\alpha > \omega \sqrt{\eps \muo}$ and $\alpha < -
\omega \sqrt{\eps \muo}$  when $\eps > 0$; and (b) $-\infty <
\alpha < \infty $ when $\eps < 0$. In geometric terms, the
wavevector components have a hyperbolic representation in
$(\alpha, \beta_{ex})$ space.
\end{itemize}

For $\mbox{Im} \lec \beta_{ex} \ric = 0 $ and $\eps_x, \eps \in
\mathbb{R}$, we find that \r{kP} reduces to
\begin{eqnarray}
\mbox{Re} \lec \#k_{ex} \ric  \. \#P (\#r)  &=& \frac{\omega^3
\muo \gamma^2 \eps^2_x }{2 \beta^2_{ex}}.
\end{eqnarray}
Hence, nonevanescent plane waves have PPV regardless of the sign
of $\gamma$ or $\eps_x$.

\subsection{Evanescent propagation}
We turn to evanescent planewave propagation in a nondissipative
medium as characterized by the inequality
\begin{equation}
\omega^2 \eps_x \muo - \alpha^2 \gamma < 0\,. \l{evan}
\end{equation}
Hence, we have $\mbox{Re} \lec \beta_{ex} \ric = 0 $.
 As in the previous subsection, we explore the cases $\gamma >
0$ and $\gamma < 0$ separately.
\begin{itemize}
\item[(i)] If $\gamma
> 0$ then the situation mirrors that which we described
earlier for hyperbolic nonevanescent propagation. That is,
evanescent propagation arises for (a) $\alpha > \omega \sqrt{\eps
\muo}$ and $\alpha < - \omega \sqrt{\eps \muo}$  when $\eps > 0$;
and (b) $-\infty < \alpha < \infty $ when $\eps < 0$. In geometric
terms, the wavevector components have a hyperbolic representation
in $(\alpha, \mbox{Im} \lec \beta_{ex} \ric )$ space.
\item[(ii)]
If $\gamma < 0$ then evanescent propagation arises provided that
$\eps > 0$, $\eps_x < 0$ and $- \omega \sqrt{\eps \muo} < \alpha <
\omega \sqrt{\eps \muo}$. In geometric terms, the wavevector
components have an elliptical representation in $(\alpha,
\mbox{Im} \lec \beta_{ex} \ric )$ space.
\end{itemize}

For $\mbox{Re} \lec \beta_{ex} \ric = 0 $ and $\eps_x, \eps \in
\mathbb{R}$,  we find that \r{kP} reduces to
\begin{eqnarray}
\mbox{Re} \lec \#k_{ex} \ric  \. \#P (\#r)  &=& \frac{ \omega
\alpha^2 \eps_x \gamma
 }{2
\le \alpha^2 \gamma - \omega^2 \muo \eps_x \ri} \,\,
 \exp \le -2
\, \mbox{Im} \lec \beta_{ex} \ric  z \ri\,.
\end{eqnarray}
Hence, evanescent plane waves have PPV if (a)  $\gamma < 0$ or (b)
$\gamma > 0$ and $\eps_x > 0$. However, negative phase velocity
(NPV) propagation arises if $\gamma > 0$ and $\eps_x < 0$.

\subsection{Illustrative examples}

Let us illustrate the geometric aspect of the dispersion relations
with some representative numerical examples.

First, suppose we consider the case $\gamma > 0$ with $\eps = 2
\epso $ and $\eps_x = 6 \epso $, where $\epso$ is the free--space
permittivity. In Figure~1 the
  real and imaginary parts of $\beta_{ex}$ are plotted against
 $\alpha$.
 The elliptical nonevanescent nature of the dispersion
 relation is clear for $- \omega \sqrt{2 \epso \muo } < \alpha < \omega \sqrt{2 \epso \muo}$, while
 the hyperbolic evanescent nature is apparent for $\alpha <
 - \omega \sqrt{2 \epso \muo}$ and $\alpha > \omega \sqrt{2 \epso \muo}$.
The elliptical/hyperbolic geometric interpretation breaks down
when dissipative mediums are considered. However, the
corresponding dispersion relations are geometrically reminiscent
of their nondissipative counterparts. This can be observed in
Figure~2  in which the graphs corresponding to Figure~1 are
displayed  for $\eps = \le 2 + i0.5 \ri \epso $ and $\eps_x = \le
6 + i 0.75 \ri \epso$.

Second, we turn to the case $\gamma < 0$ with $\eps = 2 \epso$ and
$\eps_x = -6 \epso$. The real and imaginary parts of $\beta_{ex}$
are graphed against $\alpha$ in Figure~3. The graphs mirror those
of Figure~1 but with nonevanescent and evanscent aspects
interchanged; i.e., we observe hyperbolic nonevanescent
characteristics for $\alpha <
 - \omega \sqrt{2 \epso \muo }$ and $\alpha > \omega \sqrt{2 \epso \muo}$, and elliptical evanescent
 characteristics for $- \omega \sqrt{2 \epso \muo } < \alpha < \omega \sqrt{2 \epso \muo}$. The
 corresponding graphs for
$\eps = \le  2 + i0.5 \ri \epso $ and $\eps_x = \le -6 + i 0.75
\ri \epso $ are presented in Figure~4. Notice that the shapes of
the graphs in Figures~4 and 2 are similar but not identical.

\section{Numerical conceptualization}

Although uniaxial dielectric mediums with $\gamma < 0$ do not
occur in nature (to the best of the authors' knowledge), they can
be conceptualized as metamaterials by means of homogenization.

For example, let us consider the homogenization of a composite
comprising two component materials phases, labelled as $a$ and
$b$. Both component material phases are taken to be isotropic
dielectric mediums: $\eps^a$ and $\eps^b$ denote the permittivity
scalars of phases $a$ and $b$, respectively. The component
material phases are envisioned as random distributions of
identically--oriented, spheroidal particles. The spheroidal
shape~---~which is taken to be the same for all spheroids in
component material phase $a$ and $b$~---~is parameterized via the
shape dyadic $\=U = \mbox{diag} \le U_x, U, U \ri$. That is, we
take the spheroid's principal axis to lie along the $x$ axis. The
spheroid's surface is prescribed by the vector
\begin{equation}
\#r_{\,s} (\theta, \phi) = \eta \, \=U \. \hat{\#r} (\theta,
\phi)\,,
\end{equation}
with $ \hat{\#r} $ being the radial unit vector specified by the
spherical polar coordinates $\theta$ and $\phi$. The linear
dimensions of the spheroid, as determined by the parameter $\eta$,
are  assumed to be small relative to the electromagnetic
wavelength(s).

The permittivity dyadic of the resulting homogenized composite
medium (HCM)
\begin{equation}
\=\eps^{HCM} = \mbox{diag} \le \eps^{HCM}_x, \eps^{HCM},
\eps^{HCM} \ri\,,
\end{equation}
as estimated using the Bruggeman homogenization formalism, is
provided implicitly via
\begin{equation}
f_a \, \=a^{a} + f_b \, \=a^{b}  = \=0\,, \l{Br}
\end{equation}
where $f_a$ and $f_b = 1 - f_a$ denote the respective volume
fractions of the material component phases $a$ and $b$. The
polarizability dyadics in \r{Br} are defined as
\begin{equation}
\=a^{\ell } = \le \eps^\ell \=I - \=\eps^{HCM} \ri \.\les \, \=I +
i \omega \=D \. \le \eps^\ell \=I - \=\eps^{HCM} \ri \,\ris^{-1},
\qquad (\ell = a,b),
\end{equation}
wherein the depolarization dyadic is given by the surface integral
\begin{equation}
\=D = \frac{1}{i \omega 4 \pi} \, \int^{2 \pi}_0 \, d \phi \,
\int^\pi_0 \, d \theta \, \sin \theta \, \le \frac{1}{
\hat{\#r}\.\=U^{-1}\.\=\eps^{HCM}\.\=U^{-1}\.\hat{\#r}} \ri
\=U^{-1}\.\hat{\#r} \, \hat{\#r} \. \=U^{-1}\,. \l{depol}
\end{equation}
Closed--form expressions for the depolarization dyadic for
uniaxial mediums are available in terms of hyperbolic functions
\c{Michel97}. However, we note that these exact results are not
valid for nondissipative mediums with $\gamma < 0$, and numerical
evaluation of $\=D$ has to be resorted to.

 The Jacobi iteration scheme
\begin{equation}
\=\eps^{HCM} \les \, p \, \ris = \mathcal{T} \lec \=\eps^{HCM}
\les \, p - 1 \, \ris \ric, \qquad \qquad \le \, p = 1, 2, 3,
\ldots \, \ri,
\end{equation}
where the operator $\mathcal{T}$ is defined via
\begin{eqnarray}
\mathcal{T} \lec \=\eps^{HCM} \ric &=& \Big\{ f_a  \eps^a \, \les
\, \=I + i \omega \=D \. \le \eps^a \=I - \=\eps^{HCM} \ri
\,\ris^{-1} + f_b  \eps^b \, \les \, \=I + i \omega \=D \. \le
\eps^b \=I - \=\eps^{HCM} \ri \,\ris^{-1} \Big\} \nonumber \\ &&
\. \Big\{ f_a   \, \les \, \=I + i \omega \=D \. \le \eps^a \=I -
\=\eps^{HCM} \ri \,\ris^{-1} + f_b  \, \les \, \=I + i \omega \=D
\. \le \eps^b \=I - \=\eps^{HCM} \ri \,\ris^{-1}  \Big\}^{-1},
 \nonumber \\ &&
\end{eqnarray}
may be employed to solve \r{Br} for $\=\eps^{HCM}$. Suitable
initial values for the iterative scheme are provided by
\begin{equation}
\=\eps^{HCM} \les \, 0 \, \ris = \le \, f_a \eps^a + f_b \eps^b \,
\ri \, \=I\,.
\end{equation}
For further details on  the Bruggeman homogenization formalism the
reader is referred to \c{SML00,M03} and to references therein.

Let us consider the homogenization scenario wherein material
component phase $a$ is taken to be iron  at 670 nm free--space
wavelength. Correspondingly, we take $\eps^a = \le -4.34 + i 20.5
\ri \epso $. The material component phase $b$ is assumed to be
free space; i.e., $\eps^b = \epso$. The
  shape of the component spheroids is specified by $U_x /U = 12$.
The Bruggeman estimates of the HCM permittivity parameters
$\eps^{HCM}$ and $\eps^{HCM}_x$ are plotted as functions of volume
fraction $f_a$ in Figure~5. At intermediate values of $f_a$ we see
that  $\gamma < 0$ for a substantial range of $f_a$ values.

Extensive accounts of  similar numerical homogenizations, based on
the Bruggeman formalism and more general approaches, can be found
elsewhere \c{SML00,MLW01}.

\section{Concluding remarks}

The  dispersion relations for uniaxial dielectric mediums  have
been characterized with respect to the parameter $\gamma$
\r{gamma}. For $\gamma < 0$, the dispersion relations are
hyperbolic for nondissipative mediums and hyperbolic--like for
dissipative mediums. Similarly, the dispersion relations are
elliptical for nondissipative mediums and elliptical--like for
dissipative mediums with $\gamma > 0$. Through the homogenization
of isotropic component material phases based on spheroidal
topology, we demonstrate that metamaterials with $\gamma < 0$ may
be straightforwardly conceptualized. Thus, a practical means of
achieving the exotic electromagnetic properties associated with
hyperbolic and hyperbolic--like uniaxial mediums is presented.

\newpage

\begin{figure}[!ht]
\centering \psfull \epsfig{file=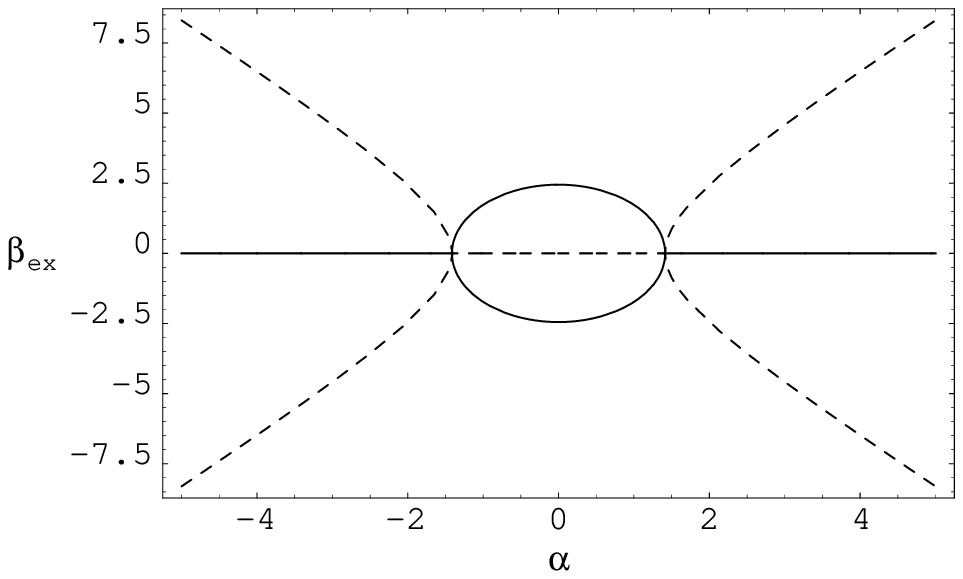,width=5.0in}
  \caption{\label{fig1} A plot of the real (solid curve) and imaginary (dashed curve)
   parts of $\beta_{ex}$ against $\alpha$ for $\eps_x = 6 \epso $ and $\eps = 2 \epso$.
The values of $\alpha$ and $\beta_{ex}$ are normalized with
respect to $\omega \sqrt{\epso \muo}$.
 }
\end{figure}

\begin{figure}[!ht]
\centering \psfull \epsfig{file=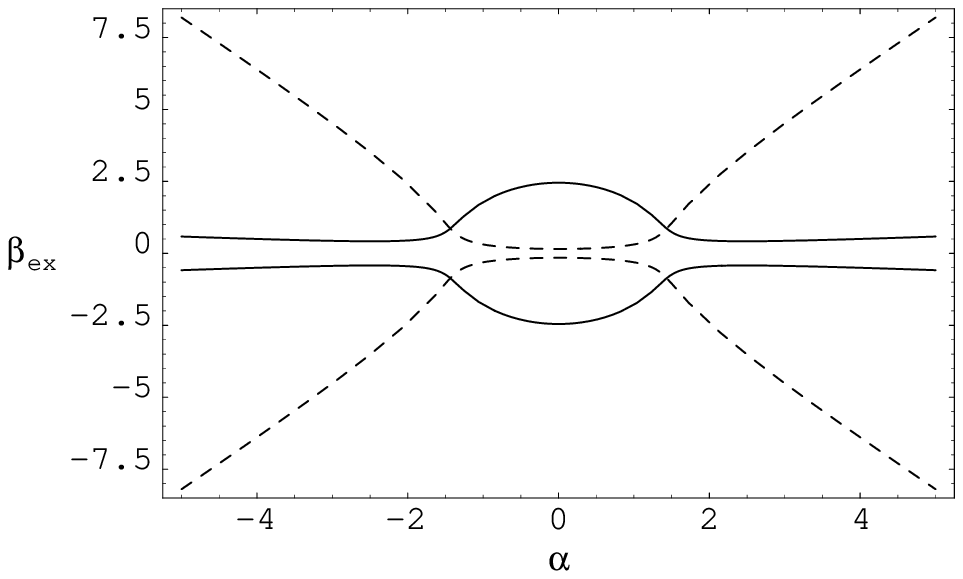,width=5.0in}
  \caption{\label{fig2} As figure~1 but for $\eps_x = \le 6 + i 0.75 \ri \epso $ and $\eps = \le
  2 + i 0.5 \ri \epso$.
 }
\end{figure}

\newpage

\begin{figure}[!ht]
\centering \psfull \epsfig{file=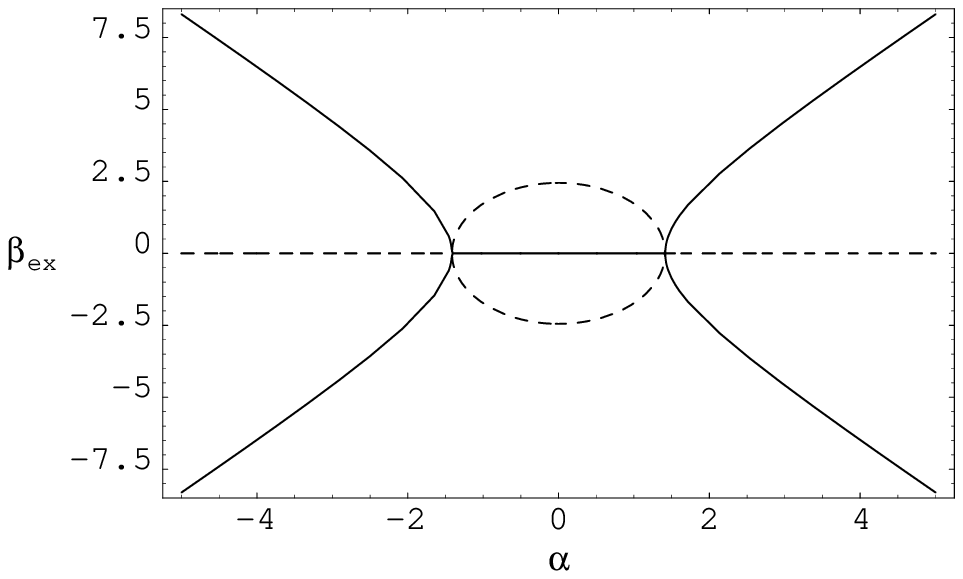,width=5.0in}
  \caption{\label{fig3} A plot of the real (solid curve) and imaginary (dashed curve)
   parts of $\beta_{ex}$ against $\alpha$ for $\eps_x = -6 \epso $ and $\eps = 2 \epso $.
The values of $\alpha$ and $\beta_{ex}$ are normalized with
respect to $\omega \sqrt{\epso \muo}$.
 }
\end{figure}

\begin{figure}[!ht]
\centering \psfull \epsfig{file=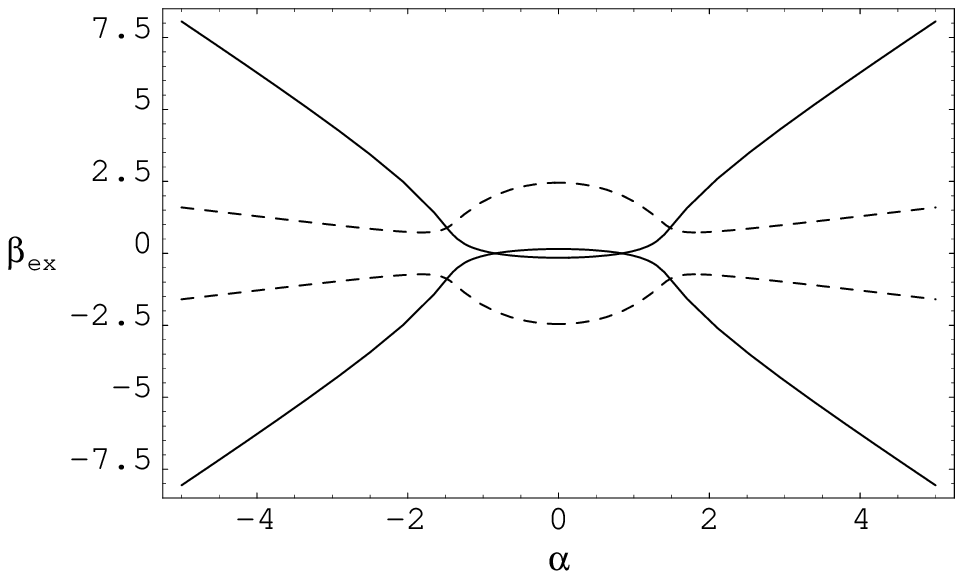,width=5.0in}
  \caption{\label{fig4} As figure~3 but for $\eps_x = \le -6 + i 0.75 \ri
  \epso  $ and $\eps = \le 2 + i 0.5 \ri \epso$.
 }
\end{figure}

\newpage

\begin{figure}[!ht]
\centering \psfull \epsfig{file=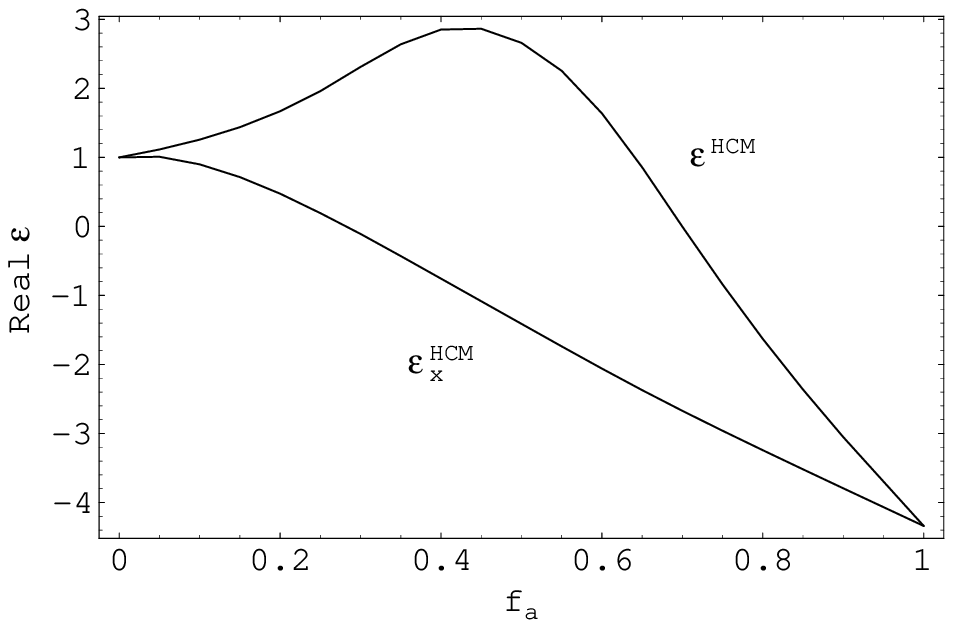,width=5.0in}
\epsfig{file=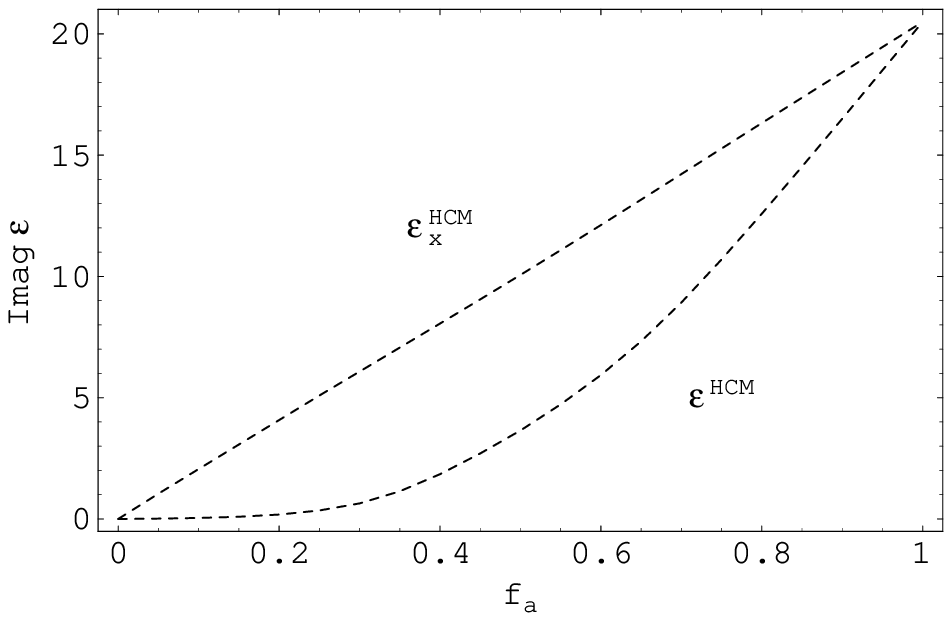,width=5.0in}
  \caption{\label{fig5} The real (above) and imaginary (below)
   parts of $\eps^{HCM}$ and $\eps^{HCM}_x$ plotted against volume
   fraction $f_a$. The permittivity values are normalized with respect to $\epso$.
   Component phase values: $\eps^a = \le
-4.34 + i 20.5 \ri \epso$ and $\eps^b = \epso$; spheroidal shape
parameters: $U_x / U = 12$.
 }
\end{figure}

\end{document}